\documentclass[aps,pra,showpacs,twocolumn]{revtex4-1}

\usepackage{bm,graphicx}
\usepackage{dsfont}
\usepackage{hyperref}
\usepackage{amssymb,amsmath,amsfonts}
\usepackage{mathrsfs}
\usepackage[none]{hyphenat}

\usepackage{color}

\usepackage[normalem]{ulem}

\usepackage[caption=false]{subfig}
\newcommand{\id}{\mathds{1}}

\DeclareMathOperator{\Tr}{Tr}

\newcommand{\ket}[1]{\left|{#1}\right\rangle}
\newcommand{\bra}[1]{\left\langle{#1}\right|}

\newcommand{\rme}{\ensuremath{\mathrm{e}}}
\newcommand{\rmi}{\ensuremath{\mathrm{i}}}

\newcommand{\hd}{\ensuremath{\mathsf{h}}}

\newcommand{\dd}{\ensuremath{\mathsf{d}}}
\newcommand{\sd}{\ensuremath{\mathsf{s}}}
\newcommand{\td}{\ensuremath{\mathsf{t}}}
\newcommand{\Sd}{\ensuremath{\mathsf{S}}}
\newcommand{\Dd}{\ensuremath{\mathsf{D}}}

\begin{document}
%
\title{Joint Schmidt-type decomposition for two bipartite pure quantum states
         }

\author{
        Christopher Eltschka$^1$ 
        and Jens Siewert$^{2,3}$  
       }
\affiliation{
$^{1}$ Institut f\"ur Theoretische Physik, Universit\"at Regensburg, D-93040 Regensburg, Germany\\
$^{2}$ Departamento de Qu\'{i}mica F\'{i}sica, Universidad del Pa\'{i}s Vasco UPV/EHU, E-48080 Bilbao, Spain\\
$^{3}$ IKERBASQUE Basque Foundation for Science, E-48013 Bilbao, Spain
}

\date{\today}
\begin{abstract}
    It is well known that the Schmidt decomposition exists for
    all pure states of a two-party quantum system.
    We demonstrate that there are two ways to obtain an 
    analogous decomposition for arbitrary rank-1 operators acting on
    states of a bipartite finite-dimensional Hilbert space. 
    These methods amount to  joint Schmidt-type 
    decompositions of two pure states where the two sets of
    coefficients and local
    bases depend on the properties of either state, however, at the
    expense of the local bases not all being orthonormal and 
    in one case the complex-valuedness of the coefficients.
    With these results we derive several generally valid purity-type
    formulae for one-party reductions of rank-1 operators, and we 
    point out relevant relations between the Schmidt decomposition
    and the Bloch representation of bipartite pure states.

\end{abstract}

\pacs{
     }

\maketitle

\section{Introduction}
%
%
The Schmidt decomposition theorem states that any pure state of
a bipartite quantum system of finite dimension
can be written as the superposition of a minimum 
number of states, where the coefficients are real and the superposed states
are tensor products of the elements of two preferred local orthonormal bases.
There are only few tools in quantum information theory comparable in
the power of the method and the ubiquity of their applicability with the 
Schmidt decomposition.
While Schmidt's original work~\cite{Schmidt1906} investigates kernels of 
integral equations,
the decomposition for finite-dimensional systems---as it is mostly applied
in quantum physics nowadays---was given by 
H.\  Everett, III~\cite{Everett1957,Ekert1995}.

There are several routes toward generalization of the method.
There is a mixed-state 
analog~\cite{Zyczkowski2006,MMWolf}, 
which is not as frequently used as the pure-state
decomposition~\cite{footnote0}, but has 
important applications, e.g., in decomposition of quantum
gates and entanglement theory~\cite{Nielsen2003,Cariello2014}.
Further, it would be desirable to have a similar method for 
multipartite states, e.g.,~\cite{Tucker1966,Carteret2000,deMoor2000,Alber2014}. 
While for three qubits this question has lead to
important results~\cite{Acin2000}, 
to date there is no generally accepted 
counterpart for the Schmidt decomposition in multipartite systems.

A third option is to ask whether there exists a simultaneous
Schmidt-type decomposition for several pure bipartite states.
In Ref.~\cite{Hiroshima2004}, the conditions for simultaneous 
applicability of a (slightly generalized) standard Schmidt decomposition
were studied;
mixtures of such jointly Schmidt-decomposable states 
(that is, Schmidt-decomposable
in one and the same pair of local bases) were then called
maximally correlated. 
However, beyond this restrictive concept nothing
seems to be known regarding simultaneous Schmidt decomposability of two or
more bipartite states. 
The reason for this is that in general the reduced states have
nonvanishing overlap and therefore it is not obvious which basis one has to
choose in the subspace of their joint support.
Our present work fills this gap by
analyzing the question of how two pairs of correlated 
local bases can be found that allow
for Schmidt-like decompositions of two arbitrary finite-dimensional 
bipartite states. We show that in general there are two options for
such decomposability, and that those generalized Schmidt bases
differ from the standard---``single-state''---Schmidt bases whenever
the reduced states have non-vanishing overlap. As a direct consequence of
these results we derive several interesting relations
for the reductions of rank-1 operators. As the Schmidt decomposition,
the existence of relations for one-party reductions, and the Bloch 
representation~\cite{Fano1957,Mahler1995,Mahler2004,Badziag2008,Kloeckl2015,Tran2015,Huber2017,ES2018} 
of bipartite quantum states are intimately related
concepts~\cite{Kloeckl2015,Huber2017,Appel2017}, 
we conclude our discussion by analyzing the most salient of these
mathematical connections.
%
%

\section{The usual Schmidt decomposition}
%
We start with a brief reminder of how the Schmidt decomposition is 
obtained following Preskill~\cite{Preskill1998}. Consider the 
generic normalized state $\ket{\psi_{AB}}$
of a bipartite Hilbert space, $\ket{\psi_{AB}}\in 
                                \mathcal{H}_A\otimes\mathcal{H}_B$,
$\langle \psi_{AB} | \psi_{AB} \rangle =1$.
We write it
with respect to some orthonormal product bases as
\begin{align}
   \ket{\psi_{AB}} = \sum_{j,k=1}^d a_{jk} \ket{j}_A\otimes\ket{k}_B
              \equiv \sum_j \ket{j}_A\otimes\ket{\tilde{j}}_B\ \ ,
\label{eq:ansatz0}
\end{align}
where the states $\ket{\tilde{j}}_B$ in general are neither normalized nor 
orthogonal. If we choose, however, for $\{\ket{j}_A\}$ the basis in which the
reduced state 
\[
    \rho_A\ =\ \Tr_B\ket{\psi_{AB}}\!\bra{\psi_{AB}}
          \ =\ \sum_{jk} \langle \tilde{k}|\tilde{j}\rangle\ket{j}_A\!\bra{k}
\]
is diagonal, the states $\{\ket{\tilde{j}}_B\}$ do become orthogonal,
and by introducing $\ket{j'}_B=\lambda_j^{-1/2}\ket{\tilde{j}}_B$
for $\lambda_j=\langle \tilde{j}|\tilde{j}\rangle > 0$ we get the 
Schmidt decomposition of $\ket{\psi_{AB}}$
\begin{align}
   \ket{\psi_{AB}}\ =\ \sum_{j,k=1}^d  \sqrt{\lambda_j}\ket{j}_A\otimes\ket{j'}_B\ \ , \ \ \sum_j\lambda_j\ =\ 1
\ \ .
\label{eq:Susual}
\end{align}
It is an immediate consequence that with this choice of local bases
also $\rho_B=\Tr_A\ket{\psi_{AB}}\!\bra{\psi_{AB}}$ is diagonal and has the
same set of nonzero eigenvalues $\{\lambda_j\}$ as $\rho_A$, so that one finds
for the purities of the local states the well-known relation
\begin{align}
 \Tr\rho_A^2\ =\ \Tr\rho_B^2\ =\ \sum_j\ \lambda_j^2\ \ .
\label{eq:purity1}
\end{align}
%

\section{Decomposition of rank-1 operators}
\subsection{Decomposition based on singular value decomposition}
%
The Schmidt decomposition Eq.~\eqref{eq:Susual} for a projector
$\ket{\psi_{AB}}\!\bra{\psi_{AB}}$ reads
\begin{align}
     \ket{\psi_{AB}}\!\bra{\psi_{AB}}\ =\ \sum_{jk} \sqrt{\lambda_j\lambda_k}
                            \ket{a_j}_A\!\bra{a_k}\otimes\ket{b_j}_B\!\bra{b_k}
\label{eq:proj}
\end{align}
with the Schmidt basis $\{ \ket{a_j}_A\otimes\ket{b_k}_B\}$. This way of writing
the decomposition imposes the following question: What happens if, in the
expression $\ket{\psi_{AB}}\!\bra{\psi_{AB}}$, 
we do not choose both sides equal,
that is, is there an analogous expansion for 
the---non-Hermitian---rank-1 operator
$\ket{\psi_{AB}}\!\bra{\phi_{AB}}$ with $\ket{\phi_{AB}}\neq\ket{\psi_{AB}}$?
An obvious idea would be to resort to the operator-level Schmidt 
decompositon, however, this way one simply retrieves the usual Schmidt 
decompsition of the individual states (cf.~Ref.~\cite{footnote0}).

In order to find a new answer we assume that 
$\ket{\psi_{AB}},\ket{\phi_{AB}}\in \mathcal{H}_A\otimes\mathcal{H}_B$ are
normalized and make an ansatz similar to what we had
in Eq.~\eqref{eq:ansatz0}
\begin{subequations}
\begin{align}
   \ket{\psi_{AB}}\ =\ &\sum_j \ket{u_j}_A|\tilde{j}^{\psi}\rangle_B
\\
   \ket{\phi_{AB}}\ =\ &\sum_k \ket{v_k}_A|\tilde{k}^{\phi}\rangle_B\ \ ,
\end{align}
\label{eq:ansatzsvd}
\end{subequations}
where $\{\ket{u_j}_A\}$, $\{\ket{v_k}_A\}$ are orthonormal bases
(from now on we will drop the tensor product signs and the lower
indices for the partitions $A$ and $B$). 
If we choose the bases  
$\{\ket{u_j}\}$, $\{\ket{v_k}\}$ such that they belong to the singular
value decomposition (SVD) of  $\Tr_B \ket{\psi}\!\bra{\phi}$ with singular
values $q_j\geqq 0$, the marginal operator for party $A$ reads
\begin{align}
      \Tr_B \ket{\psi}\!\bra{\phi} = \sum_{jk} \ket{u_j}\!\bra{v_k}
                            \langle \tilde{k}^{\phi} | \tilde{j}^{\psi}\rangle
       = \sum_j q_j \ket{u_j}\!\bra{v_j} 
\label{eq:svdB}
\end{align}
with $\langle \tilde{k}^{\phi} | \tilde{j}^{\psi}\rangle = q_j\delta_{jk}$.
That is, we find that $\{\ket{\tilde{j}^{\psi}}\}$
and $\{|\tilde{k}^{\phi}\rangle\}$ are dual bases. 
If we normalize as before
\[    \ket{\dd_j^{\psi}}\equiv \frac{ \ket{\tilde{j}^{\psi}}}{
                                      \sqrt{\mu_j^{\psi}}}
\ \ \ ,\ \ \
      \ket{\dd_k^{\phi}}\equiv \frac{ 
                                      |\tilde{k}^{\phi}\rangle}{
                                      \sqrt{\nu_k^{\phi}}}
\]
we can finally write the following for the decomposition 
of $\ket{\psi}$ and $\ket{\phi}$
\begin{subequations}
\begin{align}
\label{eq:Ssvda}
   \ket{\psi}\ =\ & \sum_j \sqrt{\mu_j^{\psi}} \ket{u_j}
                                                     |\dd_j^{\psi}\rangle\ \ ,
\\
\label{eq:Ssvdb}
   \ket{\phi}\ =\ & \sum_k \sqrt{\nu_k^{\phi}} \ket{v_k}
                                                     |\dd_k^{\phi}\rangle\ \ .
\end{align}
\label{eq:Ssvd}
\end{subequations}
One can term this the SVD-based simultaneous Schmidt-like decomposition
of $\ket{\psi}$ and $\ket{\phi}$.
We recognize the analogy of Eqs.~\eqref{eq:Ssvda} and~\eqref{eq:Ssvdb}  
with the usual 
Schmidt decomposition Eq.~\eqref{eq:Susual}. Note that, while the 
generalized ``Schmidt coefficients'' $\mu^{\psi}_j$, $\nu^{\phi}_k$ are still
real, only the bases $\{\ket{u_j}\}$, $\{\ket{v_k}\}$  are orthonormal.
The normalized bases 
$\{|\dd_j^{\psi}\rangle\}$, $\{|\dd_k^{\phi}\rangle\}$  
are dual with 
$\sqrt{\mu_j^{\psi}\nu_k^{\phi}}\langle \dd^{\psi}_j | \dd^{\phi}_k \rangle = q_j\delta_{jk}$, 
but not orthogonal. It is worthwhile noting that in general 
the orthonormal bases $\{\ket{u_j}\}$, $\{\ket{v_k}\}$ bear no special relation
with one another. Again, Eqs.~\eqref{eq:Ssvd} represent superpositions
with the minimum number of components, which equals the rank of the
reduced operator $\Tr_B\ket{\psi}\!\bra{\phi}$.
As the ``Schmidt vectors'' in Eqs.~\eqref{eq:Ssvd} are still orthogonal
we have \mbox{$\sum_j\mu_j^{\psi}=\sum_k\nu_k^{\phi}=1$}.
Moreover, there  is the condition
$\sum_j q_j\langle v_j|u_j\rangle=\langle \phi|\psi\rangle$.
Clearly, one finds an analogous decomposition with exchanged roles
of parties $A$ and $B$ by considering the reduced operator
$\Tr_A \ket{\psi}\!\bra{\phi}$ and modifying Eqs.~\eqref{eq:ansatzsvd}
correspondingly. 

The decompositions Eqs.~\eqref{eq:Ssvd} may be viewed as the result
of selecting preferred local bases for $\ket{\psi}$ and $\ket{\phi}$ depending
on the overlap of these bipartite states on party $B$ only.
This dependence  does not exist if 
$\Tr_B\ket{\psi}\!\bra{\phi}=0$. We will discuss this case below.

\subsection{Decomposition based on diagonalization}
%
Interestingly, Eqs.~\eqref{eq:Ssvd} is not the only way to obtain
a joint Schmidt-like decomposition of two pure states.
To see this, we start with an ansatz similar to Eqs.~\eqref{eq:ansatzsvd}
where we drop the assumption that the bases on party $A$ be orthonormal
\begin{subequations}
\begin{align}
   \ket{\psi}\ =\ &\sum_j \ket{x_j}\ket{\tilde{j}^{\psi}}\ \ ,
\\
   \ket{\phi}\ =\ &\sum_k \ket{y_k}|\tilde{k}^{\phi}\rangle\ \ .
\end{align}
\label{eq:ansatzdiag}
\end{subequations}
The trick that led to the decomposition was to find a diagonal form of the
reduced operator $\Tr_B\ket{\psi}\!\bra{\phi}$, which was achieved
via the singular value decomposition. There is an alternative
to this approach, namely to diagonalize $\Tr_B\ket{\psi}\!\bra{\phi}$.
Note that diagonalizability of non-Hermitian matrices is not guaranteed.
A sufficient condition is that $\Tr_B\ket{\psi}\!\bra{\phi}$ has 
the maximum number of nonzero eigenvalues, which all have
to be different~\cite{HornJohnson}.

Hence we assume that a matrix representation $\mathsf{M}$ of
$\Tr_B\ket{\psi}\!\bra{\phi}$  is similar
to a diagonal matrix $\mathsf{D}$ 
\[
\mathsf{M}\ =\ \Sd\cdot \Dd\cdot \Sd^{-1}\ \ ,
\]
where $\Sd$ is an invertible matrix. The columns of $\Sd$
are the right eigenvectors of $\mathsf{M}$, whereas the
rows of $\Sd^{-1}$ are the left eigenvectors. Correspondingly we
can write the reduced operator
\begin{align}
\Tr_B\ket{\psi}\!\bra{\phi}\ =\ \sum_j \Delta_j \rme^{\rmi\varphi_j}
                                \ket{\sd_j}\!\langle\sd^{-1}_j|\ \ ,
\end{align}
where $\{\ket{\sd_j}\}$, $|\sd^{-1}_k\rangle\}$ are 
(non-orthogonal) dual bases of $\mathcal{H}_A$, i.e.,
$\langle \sd_j | \sd^{-1}_k \rangle= \delta_{jk}$. Note 
that $\left(\langle\sd^{-1}_j|\right)^{\dagger}\neq \ket{\sd_j}$.
Moreover, we have explicitly written the phases of the eigenvalues
implying that $\Delta_j\geqq 0$.

In analogy with the discussion following Eq.~\eqref{eq:svdB}
we then find that $\ket{\tilde{j}^{\psi}}$,
$|\tilde{k}^{\phi}\rangle$ are proportional to the vectors of two
dual bases $\{\ket{\td_j}\}$, $\{|\td_k^{-1}\rangle\}$ with 
$\langle \td_j | \td^{-1}_k \rangle= \delta_{jk}$, 
so that we arrive at
\begin{subequations}
\begin{align}
   \ket{\psi}\ =\ &\sum_j \sqrt{\xi_j}\ket{\sd_j}\ket{\td_j}
\\
   \ket{\phi}\ =\ &\sum_k \sqrt{\eta_k} \rme^{-\rmi\varphi_k}
                           \ket{\sd^{-1}_k}\ket{\td^{-1}_k}
\end{align}
\label{eq:Sdiag}
\end{subequations}
with $\xi_j,\eta_k\geqq 0$ and $\sqrt{\xi_j\eta_j}=\Delta_j$.
Equations~\eqref{eq:Sdiag} represent the
second simultaneous Schmidt-like decomposition of $\ket{\psi}$ 
and $\ket{\phi}$, which now is diagonalization-based.
Here, the ``Schmidt coefficients'' have complex phases, but
there is a freedom to distribute each of the phases at will
among the two states. Interestingly, both reduced operators
$\Tr_A\ket{\psi}\!\bra{\phi}$ and $\Tr_B\ket{\psi}\!\bra{\phi}$ have
the same nonzero (now complex)
eigenvalues $\{\Delta_j\rme^{\rmi\varphi}\}$, 
analogously to the usual Schmidt decomposition of a single state.
Since the ``Schmidt vectors'' are not
orthogonal, there is no normalization condition for the $\xi_j, \eta_k$.
In contrast to the decomposition Eqs.~\eqref{eq:Ssvd}, both of the local
bases of $\ket{\psi}$ are strongly related with the corresponding 
basis in $\ket{\phi}$, however, none of them are necessarily orthogonal
or normalized. We stress again that the decomposition Eqs.~\eqref{eq:Sdiag}
may not exist, while the decomposition in Eqs.~\eqref{eq:Ssvd} 
can always be found,
hence the latter is the stronger statement.

\subsection{Remarks and special cases}
%
The obvious special case for the decompositions in Eqs.~\eqref{eq:Ssvd} and
\eqref{eq:Sdiag} is equality \mbox{$\ket{\psi}=\ket{\phi}$}. 
Here we get back the 
known result Eq.~\eqref{eq:Susual} because the 
dual bases become self-dual and therefore also orthonormal. For Hermitian
matrices the singular value decomposition coincides with diagonalization,
this ensures the usual Schmidt decomposition also for the first generalization
option. 

One might expect that also orthogonality 
\mbox{$\langle \phi | \psi \rangle = 0 $}
represents a special case, 
but as long as the reduced operators
$\Tr_A\ket{\psi}\!\bra{\phi}$, $\Tr_B\ket{\psi}\!\bra{\phi}$ do not
vanish, 
the resulting bases do not display special properties.

However, there is a case related to orthogonality
$\langle \phi | \psi \rangle = 0 $ that needs to be discussed: 
the possibility that $\Tr_A\ket{\psi}\!\bra{\phi}=0$ or/and
$\Tr_B\ket{\psi}\!\bra{\phi}=0$. Any of these conditions imply 
global orthogonality, because, e.g., 
$\langle \phi | \psi \rangle = \Tr\big( \Tr_B\ket{\psi}\!\bra{\phi} \big) = 0$.
With the latter condition, e.g., our approach 
to derive Eqs.~\eqref{eq:Ssvd} or Eqs.~\eqref{eq:Sdiag}, respectively,
does not lead to the selection of preferred bases on party $A$.
For $ \Tr_B\ket{\psi}\!\bra{\phi} = 0$
the states $\ket{\psi}$ and $\ket{\phi}$ have disjoint support 
in $\mathcal{H}_B$, 
implying also that none of them has full (usual) Schmidt rank.
Another consequence is orthogonality of the local states
$\Tr\big[\Tr_A\ket{\psi}\!\bra{\psi}\Tr_A\ket{\phi}\!\bra{\phi}\big]=0$.

In the case of disjoint support on $\mathcal{H}_B$ one can try to check 
$ \Tr_A\ket{\psi}\!\bra{\phi}$; if it is nonzero
the simultaneous Schmidt decomposition can be found as shown before, that is,
with dual bases in $\mathcal{H}_A$ and singular value decomposition
(or diagonalization) in $\mathcal{H}_B$.
If, however,  also $ \Tr_A\ket{\psi}\!\bra{\phi}=\Tr_B\ket{\psi}\!\bra{\phi}=0$,
the states have disjoint support on the entire composite Hilbert space 
$\mathcal{H}_A\otimes\mathcal{H}_B$, and it is
not possible (but also not necessary) to select preferred local bases 
whose properties depend on both $\ket{\psi}$ and $\ket{\phi}$. It suffices then
to diagonalize the local states of $\ket{\psi}$ and $\ket{\phi}$ separately and 
to use their standard Schmidt decomposition.

\section{Generalized purity relations}
%
As an immediate application of the decomposition 
Eqs.~\eqref{eq:Ssvd}, \eqref{eq:Sdiag} we derive several formulae
that may be regarded as the generalizations of the
purity relation Eq.~\eqref{eq:purity1}.

Consider first the squares of the reduced rank-1 operators. 
By using Eq.~\eqref{eq:svdB}
we find $\Tr\big(\Tr_B\ket{\psi}\!\bra{\phi}\big)^2=
         \sum_{jk} q_jq_k \langle v_j|u_k\rangle\langle v_k|u_j\rangle$.
On the other hand, $\Tr_A\ket{\psi}\!\bra{\phi}= 
                  \sum_{jk} (\mu^{\psi}_j\nu^{\phi}_k)^{1/2}\langle
                  v_k|u_j\rangle|\dd_j^{\psi}\rangle\!\langle\dd_k^{\phi}|$.
Because of 
$(\mu^{\psi}_j\nu^{\phi}_k)^{1/2}
 \langle\dd_k^{\phi}| \dd_j^{\psi}\rangle = q_j\delta_{jk}$ it follows that
\begin{align*}
   \Tr\big(\Tr_A\ket{\psi}\!\bra{\phi}\big)^2= & \sum_{jklm}
                    \sqrt{\mu^{\psi}_j\nu^{\phi}_k} 
                     \sqrt{\mu^{\psi}_l\nu^{\phi}_m}\ \ \times &
\\        &  \times\
\langle v_k|u_j\rangle\langle v_m|u_l\rangle 
\langle \dd_k^{\phi}|\dd_l^{\psi}\rangle\langle \dd_m^{\phi}|\dd_j^{\psi}\rangle
& \hfill
\\      = & \
         \sum_{jk} q_jq_k \langle v_j|u_k\rangle\langle v_k|u_j\rangle\ \ ,
\end{align*}
so that we conclude
\begin{align}
            \Tr\big(\Tr_A\ket{\psi}\!\bra{\phi}\big)^2
   \ =\
            \Tr\big(\Tr_B\ket{\psi}\!\bra{\phi}\big)^2\ \ .
\label{eq:puritysq}
\end{align}
If the reduced operators are diagonalizable Eq.~\eqref{eq:puritysq}
follows practically without calculation, because  
$\Tr_A\ket{\psi}\!\bra{\phi}$ and $\Tr_B\ket{\psi}\!\bra{\phi}$ have the 
same (in general complex) eigenvalues.
 
Alternatively, one can read Eq.~\eqref{eq:purity1} as Hilbert-Schmidt
scalar products. Then, we derive in a completely analogous manner
the second generalized purity relation
\begin{align}
            \Tr\bigg[\Tr_B\ket{\psi}\!\bra{\phi} & \Tr_B\ket{\phi}\!\bra{\psi}
               \bigg]
    \ =\  \sum_j q_j^2 \ =
\nonumber\\
      =\ & \Tr\bigg[\Tr_A\ket{\psi}\!\bra{\psi}\Tr_A\ket{\phi}\!\bra{\phi}
               \bigg] \ .
\label{eq:purityadj}
\end{align}
%
We can go one step further and turn
Eqs.~\eqref{eq:puritysq}, \eqref{eq:purityadj} into a single equality
linking the reductions of four 
bipartite states $\psi,\phi,\chi,\zeta\in\mathcal{H}_A\otimes \mathcal{H}_B$
\begin{align}
            \Tr\bigg[\Tr_A\ket{\psi}\!\bra{\chi}\Tr_A\ket{\phi}\!\bra{\zeta}
               \bigg]
    =
            \Tr\bigg[\Tr_B\ket{\psi}\!\bra{\zeta}\Tr_B\ket{\phi}\!\bra{\chi}
               \bigg] \ .
\label{eq:purity4}
\end{align}
The relations Eq.~\eqref{eq:puritysq}--\eqref{eq:purity4}
constitute the second central result of our article.
They are directly connected with the Bloch representation
of quantum states; therefore they are highly useful 
in calculations within that formalism, as will be demonstrated
in forthcoming work.
We will highlight some of the links to the Bloch representation
in the last part of our discussion.

\section{Schmidt decomposition and Bloch representation}
%
Our main point in generalizing the Schmidt decomposition was to
consider rank-1 operators such as $\ket{\psi}\!\bra{\phi}$ and
$\ket{\psi}\!\bra{\psi}$ rather than state vectors like $\ket{\psi}$.
Moreover, we have seen in the preceding sections that a discussion of 
the Schmidt decomposition (both the usual and the joint version) is closely 
related to analyzing the properties of the marginals. Thus one is led
to associate the entire discussion with yet another topic where operators
and their marginals are intrinsically tied together, namely the Bloch
representation of quantum states. This link is rarely emphasized,
therefore we use this section to work out some of its details.
We will restrict this analysis mainly to properties of the usual 
(single-state) Schmidt decomposition, cf.~Sec.~II, however, some of
the results we derived earlier will turn out to be useful.
In our considerations we will assume Hilbert spaces of
equal dimension, $\dim\mathcal{H}_A=\dim\mathcal{H}_B=d$ (this can always
be done by extending the Hilbert space of lower dimension), because it
makes the expressions more transparent.

The Bloch representation of bipartite states 
is defined as follows (cf., e.g.,~\cite{Fano1957,Mahler1995,Mahler2004,Badziag2008,Kloeckl2015,Tran2015,Huber2017,ES2018}):
Given an orthonormal basis of trace-free Hermitian matrices $\{\hd_j\}$
(with normalization $\Tr\left(\hd_j \hd_k\right)= d\, \delta_{jk}$ and
$\hd_0\equiv\id$)
we can expand any density operator 
$\rho$ acting on $\mathcal{H}_A\otimes\mathcal{H}_B$
\begin{align}
 \rho\ =\ & \frac{1}{d^2}\bigg[ \left(\Tr\rho\right)\id\otimes\id   + 
                       \sum_{j=1}^{d^2-1} r_{j0} \hd_j\otimes \id  + \bigg.
\nonumber\\
               &  \bigg.   + \sum_{k=1}^{d^2-1} r_{0k} \id\otimes \hd_k +
                       \sum_{l,m=1}^{d^2-1} r_{lm}\hd_l\otimes\hd_m
                           \bigg]\ \ .
\label{eq:Blochdef}
\end{align}
With the simplifying choice of Hermitian matrices $\hd_j$
the coefficients $r_{j0}$, $r_{0k}$ and $r_{lm}$ are real.
We call the sum of terms with one summation index ``1-sector'',
and the one with two indices ``2-sector''. The first term
on the right-hand side of Eq.~\eqref{eq:Blochdef} can 
be denoted the ``0-sector'', accordingly.
The purity
condition for the state $\rho$ translates into 
the Bloch vector length as the sum 
of sector lengths~\cite{Kloeckl2015,Tran2015,Huber2017,ES2018}
\begin{align}
  d^2  \Tr\rho^2 =  \left(\Tr\rho\right)^2 + \sum_{j=1}^{d^2-1} r_{j0}^2
                 + \sum_{k=1}^{d^2-1}r_{0k}^2 + \sum_{l,m=1}^{d^2-1} r_{lm}^2
                                  \ .
\label{eq:seclengths}
\end{align}
Each of the 
sums in Eq.~\eqref{eq:seclengths} is
invariant under local unitary transformations~\cite{footnote1}.
The purity condition~\eqref{eq:purity1} for the reduced states 
corresponds to the 
equality of the normalized 1-sector lengths 
$\sum_j r_{j0}^2=\sum_k r_{0k}^2$. 
Moreover, we note that for
normalized states  $\sum_{j,k=0}^{d^2-1} r_{jk}^2\leqq d^2$.

Consider now the Bloch representation of the rank-1 operator 
$\ket{\psi}\!\bra{\phi}$,
\begin{align*}
\ket{\psi}\!\bra{\phi} \ =\ 
        & \frac{1}{d^2} \sum_{lm} x_{lm}\hd_l\otimes\hd_m \ \ .
\label{eq:Blochoff}
\end{align*}
The coefficients $x_{jk}$ in general are complex.
For normalized $\ket{\psi}$, $\ket{\phi}$ it follows that
    $\Tr\big[\left(\ket{\psi}\!\bra{\phi}\right)^{\dagger}
                   \ket{\psi}\!\bra{\phi}
       \big] = 1$, hence the total length of this rank-1 operator is
\begin{align}
    \sum_{j,k=0}^{d^2-1} |x_{jk}|^2\ =\ d^2\ \ .
\end{align}
With our discussion above and Eq.~\eqref{eq:purityadj}
we find that in general $\sum_j|x_{j0}|^2 \neq \sum_k|x_{0k}|^2$.
For example, it is well possible that 
$\sum_{j=0}^{d^2-1}|x_{j0}|^2=0$ while still 
$\sum_{k=0}^{d^2-1}|x_{0k}|^2\neq 0$.
As we discussed, $\Tr_B\ket{\psi}\!\bra{\phi}=0$ implies $\langle \psi|\phi
\rangle=0$, hence in this case the Bloch representation consists only
of the 1-sector corresponding to party $B$, and the 2-sector.
If both $\Tr_A\ket{\psi}\!\bra{\phi}=\Tr_B\ket{\psi}\!\bra{\phi}=0$,
only the 2-sector has non-vanishing components. 

The Schmidt decomposition in a way captures the essence
of superposition for bipartite states. On the other hand, 
for density 
operators---as described by the Bloch representation---superposition
is not a concept as obvious as for state vectors. Let us therefore 
elaborate further on the properties of superpositions in the
Bloch formalism. 
Consider the 
superposition of several normalized orthogonal
states $\ket{\phi_j}\in\mathcal{H}_A\otimes\mathcal{H}_B$, $1\leqq j \leqq d^2$
%
%
%
\begin{align}
 \ket{\Psi}\ =\ \sum_j a_j \ket{\phi_j}\ \ .
\label{eq:sup}
\end{align}
The projector onto $\ket{\Psi}$ naturally splits up into a diagonal and an
offdiagonal part,
$\ket{\Psi}\!\bra{\Psi}=\mathsf{diag}+\mathsf{offdiag}$,
\begin{subequations}
\begin{align}
     \mathsf{diag}\ = \ & \sum_j |a_j|^2 \ket{\phi_j}\!\bra{\phi_j} 
                        \ ,
\label{eq:diag-offdiagA}
\\
\mathsf{offdiag}\ = \ & \sum_{j<k} a_j a_k^* \ket{\phi_j}\!\bra{\phi_k}
                  + a_j^* a_k \ket{\phi_k}\!\bra{\phi_j} \ \ ,
\end{align}
\label{eq:diag-offdiag}
\end{subequations}
which are orthogonal 
$\Tr\big(\mathsf{diag}^{\dagger}\mathsf{offdiag}\big) = 0 $.
Hence, for the Bloch vector length of $\ket{\Psi}\!\bra{\Psi}$ divided
by $d^2$ we
have $1= \Tr\big(\mathsf{diag}^{\dagger}\mathsf{diag}\big)+
\Tr\big(\mathsf{offdiag}^{\dagger}\mathsf{offdiag}\big)=
\sum_j |a_j|^4+2\sum_{j<k} |a_j|^2 |a_k|^2$. 
%

For bipartite systems one commonly chooses $\ket{\phi_j}$ 
in Eq.~\eqref{eq:sup}
as tensor products of local basis states in order to distinguish local
from nonlocal physics. 
It is then convenient to use two summation indices 
$\ket{\Psi}=\sum_{kl} a_{kl}\ket{e_kf_l}$ with local 
orthonormal bases $\{\ket{e_k}\}$,
$\{\ket{f_l}\}$, $(k,l=1\ldots d)$.
The matrices $\{\hd_j\}$ (e.g.,
the generalized Gell-Mann matrices~\cite{ES2018}) 
refer to the same local bases.
Among all the possible local bases the (usual) Schmidt basis of 
$\ket{\Psi}$ is peculiar: $\ket{e_kf_k}$  
become the Schmidt vectors and 
$a_{kl}\longrightarrow \sqrt{\lambda_k}\ \delta_{kl}$
the Schmidt coefficients. 
As then 
$\Tr_A\ket{\Psi}\!\bra{\Psi}$ and $\Tr_B\ket{\Psi}\!\bra{\Psi}$ are
diagonal, the Bloch representation of $\ket{\Psi}\!\bra{\Psi}$ contains
only diagonal matrix terms in the 1-sector. Remarkably, the length
$\Tr\big(\mathsf{offdiag}^{\dagger}\mathsf{offdiag}\big)=
2\sum_{j<k}\lambda_j\lambda_k$ equals
half the squared concurrence of $\ket{\Psi}$~\cite{Rungta2001}. 

In Eq.~\eqref{eq:diag-offdiag} we can recognize the 
importance of the second generalized purity relation Eq.~\eqref{eq:purityadj}:
If one wants to describe the parts of a superposition in terms of the
Bloch vector coefficients of the superposed states $\ket{\phi_j}\!\bra{\phi_j}$
this is obvious for the diagonal part. In contrast, it is not clear
whether there is any simple relation between the rank-1 operators in 
$\mathsf{offidag}$, 
cf.~Eq.~\eqref{eq:diag-offdiagA}, and the Bloch coefficients.
Here, Eq.~\eqref{eq:purityadj} provides an answer.

We may ask what the contributions of
$\mathsf{diag}$ and $\mathsf{offdiag}$  to the sectors of the Bloch
representation are. Both 1- and 2-sector lengths 
are invariant under local
unitaries, therefore a local basis change leads to a redistribution of
the parts that $\mathsf{diag}$ and $\mathsf{offdiag}$ contribute 
to the 1-sector or the 2-sector,
respectively. The Schmidt decomposition is special, because 
$\Tr_A\left(\ket{e_kf_k}\!\bra{e_lf_l}\right)=
 \Tr_B\left(\ket{e_kf_k}\!\bra{e_lf_l}\right)=0$
(for $k\neq l$), that is,
$\mathsf{offdiag}$ does not contribute to the 1-sector at all, while
(trivially) the $\mathsf{diag}$ contribution to the 1-sector is maximum.
However, a nontrivial fact is that the $\mathsf{diag}$ contribution
to the 2-sector  also has its maximum in the Schmidt basis (for the proof, see
Appendix),
\begin{subequations}
\begin{align}
     \Tr\big[(\text{2-sector})\cdot \mathsf{diag}\big]\ 
&    \overset{\longrightarrow}{
     \text{\tiny Schmidt\ decomp.}}\ \max\ \ ,
\\
     \Tr\big[(\text{2-sector})\cdot \mathsf{offdiag}\big]\ 
&    \overset{\longrightarrow}{
     \text{\tiny Schmidt\ decomp.}}\ \min\ \ ,
\label{eq:2sector-b}
\end{align}
\label{eq:2sector}
\end{subequations}
%
where ``2-sector''\ $= \sum_{jk} r_{jk} \hd_j\otimes\hd_k$,
as defined before in Eq.~\eqref{eq:Blochdef},
so that the entire $\mathsf{diag}$ length is maximum
in the Schmidt basis, whereas the $\mathsf{offdiag}$ length is 
minimum and equals half of the squared concurrence, cf.~Fig~1.
%
\begin{figure}[thb]
  \centering
  \includegraphics[width=.88\linewidth]{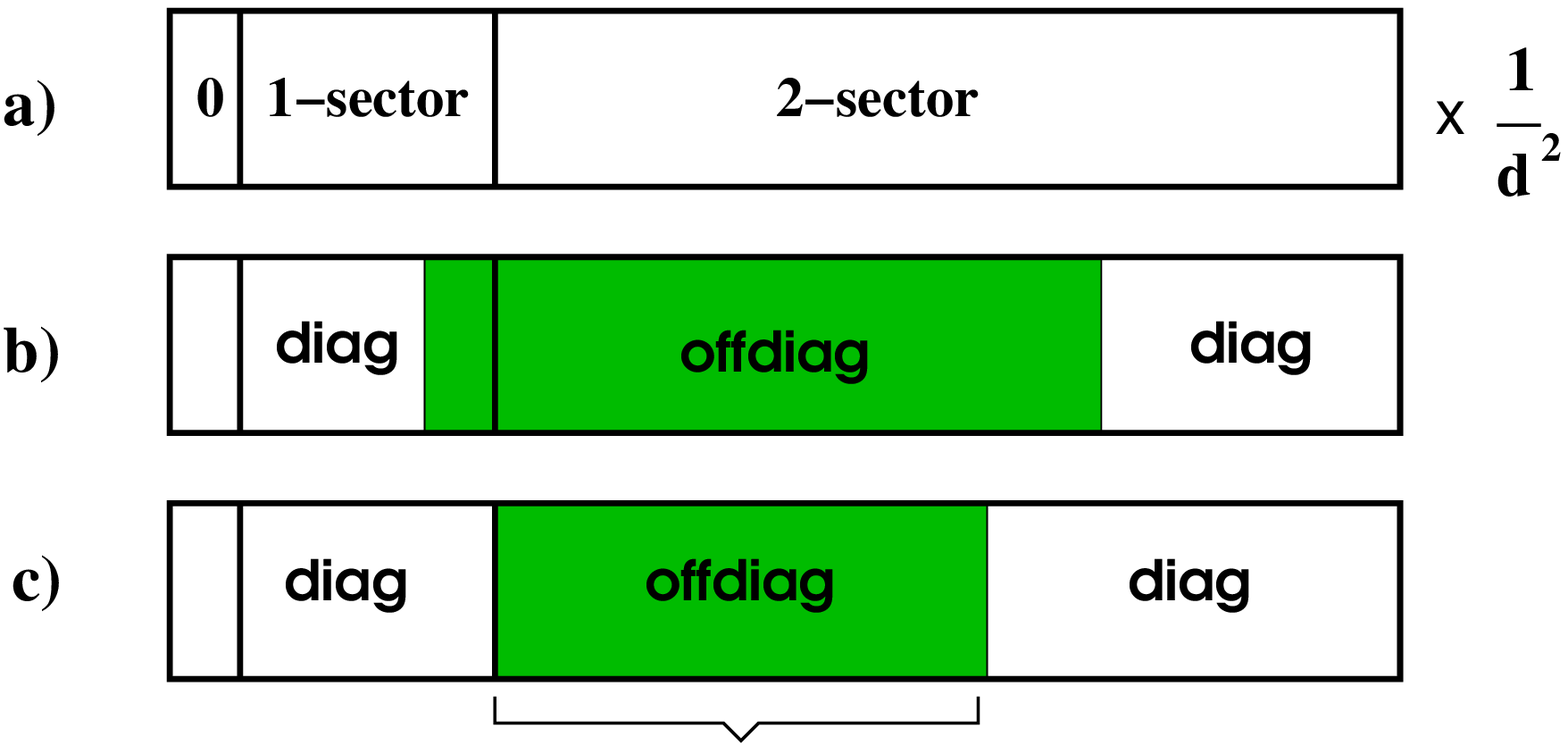}
\mbox{ }\hspace{.4\linewidth}$\frac{1}{2}C^2(\Psi)$
  \caption{Schematic of the $\mathsf{diag}$/$\mathsf{offdiag}$
           contributions to the 1-sector/2-sector of $\ket{\Psi}\!\bra{\Psi}$
           for different choices of local bases. The 0-sector per definition
           belongs to $\mathsf{diag}$.\\ 
           a) The lengths of 0-sector, 1-sector and 2-sector are invariant
              under local basis changes.
           b) Generic local bases: $\mathsf{diag}$ and
              $\mathsf{offdiag}$ contribute to both the 1-sector and the 
              2-sector.
           c) Schmidt basis: $\mathsf{offdiag}$ does not contribute
              to the 1-sector and has minimum length, which equals
              half the squared concurrence $\frac{1}{2}C^2(\Psi)$.
              The $\mathsf{diag}$ contribution to 
              the 2-sector is maximum.
              Correspondingly, also the total length of $\mathsf{diag}$
              is maximum. The total length of all contributions
              in this figure is 1 [note that the sum of sector
              lengths, Eq.~\eqref{eq:seclengths}, 
              per definition equals $d^2$].
    }
  \label{fig:fig1}
\end{figure}

%
This illustrates an archetypical situation for the 
Bloch formalism: In a parametrically interesting regime (here, the special
choice of the Schmidt bases), various relevant quantities
assume extreme values, such as the total $\mathsf{diag}$ length and
the 1-sector $\mathsf{diag}$ contribution. However, intriguingly, also
the difference of these maximum values, the 2-sector $\mathsf{diag}$ 
part is {\em maximum}. We mention that this fact was observed 
independently by M.\ Huber~\cite{Marcus}.
%

\section{Conclusions}
%
We have devised two ways to obtain a simultaneous
Schmidt-type decomposition of two arbitrary bipartite pure states in
finite dimensions, Eqs.~\eqref{eq:Ssvd} and Eqs.~\eqref{eq:Sdiag},
based on singular value decomposition 
of the reduced rank-1 operator on the one hand,
and on diagonalization thereof, on the other.
The corresponding ``Schmidt bases'' depend on the overlap of the reduced
states in the local Hilbert spaces; consequently, if there is no overlap
one can simply use the standard Schmidt decomposition. 
It is surprising that the simultaneous Schmidt decompositions 
maintain the simplicity and many of the important properties of 
the single-state Schmidt
decomposition, such as a minimum number of (real) coefficients that equals 
the rank of the reduced operator.
As an immediate consequence from these decompositions we have derived
several interesting purity-type relations for the reductions of 
bipartite pure states, Eq.~\eqref{eq:puritysq}-\eqref{eq:purity4}.
Moreover we used these results to analyze the mathematical relations
between the Bloch representation and the Schmidt decomposition. 
To this end, we introduced the diagonal and offdiagonal operator
parts $\mathsf{diag}$, $\mathsf{offdiag}$
of a projector which make explicit the extreme properties of 
the Schmidt bases regarding their contribution to the sector
lengths of the Bloch vector. 

\section{Acknowledgements}
This work was funded by the German Research Foundation Project
EL710/2-1, by Basque Government grant IT986-16 and
MCIU/FEDER/UE grant  PGC2018-101355-B-I00 (J.S.). 
The authors acknowledge illuminating discussions with Marcus Huber,
and Klaus Richter's support for this project. 
%
%

%

\renewcommand{\thesection}{\Alph{section}}
\renewcommand{\theequation}{\thesection\arabic{equation}}
\setcounter{equation}{0}
\section*{Appendix}
\setcounter{section}{1}
Here we prove Eq.~\eqref{eq:2sector} in the main text,
i.e., the statement that the contribution of $\mathsf{diag}$ to
the 2-sector of the Bloch representation of a pure bipartite
state $\ket{\Psi}=\sum_{jk} a_{jk}\ket{jk}$, $1\leqq j,k\leqq d$
assumes its maximum for the Schmidt basis and, correspondingly,
the $\mathsf{offdiag}$ has its minimum contribution to the 2-sector,
where $\{\ket{j}\}$,
$\{\ket{k}\}$ are orthonormal bases of $\mathcal{H}_A$ and $\mathcal{H}_B$,
respectively, and $\sum_{jk}|a_{jk}|^2=1$. What we will show, in fact,
is the second statement, Eq.~\eqref{eq:2sector-b}. The maximum of 
the $\mathsf{diag}$ contribution to the 2-sector follows immediately
by recalling that the total length of the 2-sector is invariant under
local unitaries. Note that it would not be sufficient for the 
proof of Eq.~\eqref{eq:2sector} to show
that the full $\mathsf{offdiag}$ part is minimum for the Schmidt basis.

Consider the reduced state of party $A$
\begin{align}
      \rho_A\ =\ \Tr_B\ket{\Psi}\!\bra{\Psi}\ =\
            \sum_{jk}\left( \sum_l a_{jl}a^*_{kl}\right)\ket{j}\!\bra{k}
\end{align}
with diagonal elements 
\begin{align}
       h_j\ =\ \sum_l |a_{jl}|^2\ \ .
\label{eq:diagh}
\end{align}
Assume that the Schmidt basis of $\ket{\Psi}$ on party $A$ is
\begin{align}
      \ket{e_m} \ =\ \sum_n U_{mn}\ket{n}\ \ ,
\end{align}
where $U_{mn}$ is a unitary matrix. In this basis $\rho_A$ is
diagonal,
\begin{align}
    \rho_A\ =\ \sum_j \lambda_j\ket{e_j}\!\bra{e_j}
\end{align}
with  the Schmidt coefficients $\lambda_j$ of $\ket{\Psi}$.
One readily obtains
\begin{align} 
     h_j\ =\ \sum_k |U_{jk}|^2 \lambda_k\ \equiv\ \sum_k M_{jk}\lambda_k\ \ ,
\end{align} 
where the matrix $M$ is doubly stochastic because of $|U_{jk}|^2\geqq 0$ and
$\sum_k M_{jk}=\sum_j M_{jk}=1$. According to the Hardy-Littlewood-P\'olya
theorem~\cite{Steele} this means that the vector of Schmidt coefficients,
$\lambda$, majorizes the vector $h=M\lambda$ of diagonal 
entries of $\rho_A$ written 
in the basis $\{\ket{j}\}$,
\begin{align}
             \lambda\ \succ\ h\ \ . 
\label{eq:maj}
\end{align}
For a Schur concave function $f(h_1,h_2,\ldots,h_d)$, Eq.~\eqref{eq:maj}
implies
\begin{align}
     f(\lambda_1,\lambda_2,\ldots,\lambda_d)\ \leqq\ 
                   f(h_1,h_2,\ldots,h_d)\ \ .
\label{eq:concave}
\end{align}
Now it is known that the elementary symmetric functions are Schur
concave~\cite{Steele}. Here we are interested in the 2nd elementary symmetric
function
\begin{align}
         S_2(h_1,h_2,\ldots,h_d) \ =\ \sum_{j<k} h_j h_k\ =\ S_2(h)\ \ .
\end{align}
If we substitute Eq.~\eqref{eq:diagh} and apply Eq.~\eqref{eq:concave} we obtain
\begin{align}
    2 \sum_{j<l} h_j h_l\ =\ & \sum_{j\neq l} \sum_{km} |a_{jk}|^2 |a_{lm}|^2
\nonumber\\
     \ \geqq\  & \sum_{j\neq l}\lambda_j\lambda_l
\label{eq:preoff}
\\
\ =\ & \frac{1}{2}C^2(\Psi)
\ \ .
\nonumber
\end{align}
Note that the summation in the first line of this equation has to be
understood as 
\[
\sum_{k,l,m=1}^d\sum_{\scriptsize
                     \begin{array}{cc}
                              j=1\\
                              j\neq l
                     \end{array}}^d\ \ ,
\]
that is, 
the three indices $k,l,m$ are summed without restriction, while the fourth
index $j$ must be different from $l$.

Finally we symmetrize Eq.~\eqref{eq:preoff} with respect to the parties
$A$ and $B$ (the entire discussion up to this point considered party $A$;
but since the vector of eigenvalues is the same for $\rho_B$, it is 
equally valid for party $B$)
\begin{align}
    2 \sum_{j<l} h_j h_l = & \frac{1}{2}\left(
                    \sum_{j\neq l,km}  |a_{jk}|^2 |a_{lm}|^2\ +\
                    \sum_{k\neq m,jl}  |a_{jk}|^2 |a_{lm}|^2
                                          \right)
\nonumber\\
      \geqq  & \ \sum_{j\neq l}\lambda_j\lambda_l
\label{eq:preoffsymm}
\\
 = &\ \frac{1}{2}C^2(\Psi)
\ \ .
\end{align}

In order to finish the proof of Eq.~\eqref{eq:2sector-b}
we need to explain the relation of the expression in the first line
of Eq.~\eqref{eq:preoffsymm} with the $\mathsf{offdiag}$ part in
the 2-sector of $\ket{\Psi}\!\bra{\Psi}$. The complete
$\mathsf{offdiag}$ part is given by
\begin{align}
    \mathsf{offdiag}(\Psi)\ =\ \sum_{(jk)\neq(lm)} a_{jk}a^*_{lm}
                                  \ket{jk}\!\bra{lm}\ \ ,
\end{align}
and, hence, its length
\begin{align}
    \Tr\big[\mathsf{offdiag}(\Psi)^{\dagger}\mathsf{offdiag}(\Psi)\big]
        = \sum_{(jk)\neq(lm)} |a_{jk}|^2 |a_{lm}|^2\  .
\label{eq:offsum-all}
\end{align}
In this sum, the index pair $(jk)$ must not coincide with the
pair $(lm)$. This is achieved by 
\begin{align}
       \sum_{k,l,m=1}^d\sum^d_{\scriptsize
                               \begin{array}{cc}j=1\\ 
                                  j\neq l\ \text{if}\ k=m
                               \end{array}}\ =\
       \sum_{k,l,m=1}^d\sum^d_{\scriptsize
                               \begin{array}{cc}j=1\\ 
                                  j\neq l 
                               \end{array}}
          + 
       \sum_{j,k,l=1}^d\sum^d_{\scriptsize
                               \begin{array}{cc}m=1\\ 
                                  m\neq k 
                               \end{array}} \delta_{jl}
\label{eq:sum-rule}
\ \ .
\end{align}
In order to obtain the
length $\ell_{\text{off,2-sec}}^2$ of the $\mathsf{offdiag}$
contribution to the 2-sector we need to subtract the $\mathsf{offdiag}$
parts of the 1-sector
\begin{align}
\ell_{\text{off,2-sec}}^2\ 
        = \ & \sum_{(jk)\neq(lm)} |a_{jk}|^2 |a_{lm}|^2\  -
\nonumber\\
            & -\ \frac{1}{d}\sum_{j\neq l,k} |a_{jk}|^2|a_{lk}|^2\ -
\nonumber\\
            & -\ \frac{1}{d}\sum_{j,k\neq l} |a_{jk}|^2|a_{jl}|^2\ \ .
\label{eq:offsum-2sec}
\end{align}
%
By symmetrizing the summation rule~\eqref{eq:sum-rule} with respect
to parties $A$ and $B$ and applying it to Eq.~\eqref{eq:offsum-2sec}
we find
\begin{align}
\ell_{\text{off,2-sec}}^2\ 
        =  & \frac{1}{2}\left(
                 \sum_{j\neq l,km} |a_{jk}|^2 |a_{lm}|^2  +
                 \sum_{k\neq m,jl} |a_{jk}|^2 |a_{lm}|^2  
                         \right) +
\nonumber\\
            & +\ \left(\frac{1}{2}-\frac{1}{d}
                 \right)\sum_{j\neq l,k} |a_{jk}|^2|a_{lk}|^2\ +
\nonumber\\
            & +\ \left(\frac{1}{2}-\frac{1}{d}
                 \right)\sum_{j,k\neq l} |a_{jk}|^2|a_{jl}|^2\ \ .
\label{eq:offsum-2sec2}
\end{align}
%
Thus, since $d\geqq 2$, the sum in Eq.~\eqref{eq:offsum-2sec2} contains more 
(non-negative) terms than the one in Eq.~\eqref{eq:preoffsymm}, so that
\begin{align}
  \ell_{\text{off,2-sec}}
  \geqq
    2 \sum_{j<l} h_j h_l \geqq 2\sum_{j\neq l} \lambda_j\lambda_l 
 =  \frac{1}{2}C^2(\Psi) \ .
\end{align}
This inequality is tight, as 
 $\ell_{\text{off,2-sec}}=\frac{1}{2}C^2(\Psi)$
in the Schmidt basis.
Thus  our proof is complete.\hfill $\square$
\vspace*{4\baselineskip}



\begin{thebibliography}{99}
\bibitem{footnote0}
In the case of pure states, the operator Schmidt decomposition
essentially retrieves the pure-state decomposition; it is not
difficult to see this by considering Eq.~\eqref{eq:proj} and viewing
$(jk)$ as a single two-digit index that runs from 1 through $d_A d_B$,
the dimension of an appropriate operator basis.
\end{thebibliography}
\end{document}